# Multipartite Entanglement in a Discrete Magnetic Bands Magnetic Lattice

A. Abdelrahman[1*], P. Hannaford[2], M. Vasiliev[1] and K. Alameh[1]

[1] *Electron Science Research Institute, Edith Cowan University, 270 Joondalup Drive, Joondalup WA 6027 Australia.*
[2] *Centre for Atom Optics and Ultrafast Spectroscopy, and ARC Centre of Excellence for Quantum Atom Optics, Swinburne University of Technology, Melbourne, Australia 3122*
\* a.abdelrahman@ecu.edu.au

## Abstract

Using magneto-optical materials, an asymmetric multi-quantum state magnetic lattice is proposed to host an entangled multipartite system formed by using ultracold atoms in a Bose Einstein Condensate (BEC). The discrete magnetic bands magnetic lattice is devised to locate a controllable long-range entanglement of many qubits separated in space. The confinement of the coupled magnetic quantum well (CMQW) system may significantly improve the condition for the long-range entanglement of the multipartite that to be used in quantum information processors.

## 1. INTRODUCTION

Long-range entanglement provides coherent coupling between many qubit quantum systems leading to the required long time of coherence to achieve a reasonable number of quantum computing operations [1,2,4]. It is also considered as one of the big experimental and theoretical challenges in the field of quantum computing and quantum information processing and communication, where several proposals have been considered to demonstrate the possibility of long-range entanglement of many qubits [5]. Most of these approaches exhibit mixed states due to the coupling of the qubit pure quantum state to thermal states. Scaling and quantifying the degree of entanglement (entropy of entanglement) for a mixed state is not uniquely identified. However there are some theoretical frames considered to partially quantify the entanglement of composite quantum states, such as the entanglement distillation method [6]. The crucial point in these approaches is to achieve the long-range entanglement with the lowest possible dimensions of the complex Hilbert space of the system, specifically the requirement for individual local control. Based on this dynamical fact we propose a quantum device that is capable of providing a suitable environment for a long-range entangled multipartite system (many qubits) created using trapped ultracold atoms in a discrete magnetic bands magnetic lattice. Magnetically trapped ultracold atoms represent stable harmonic quantum systems. Atoms are decelerated by means of laser cooling and magnetic trapping and pumped into the so-called low magnetic field seeking state in which the ultracold atoms tend to settle in periodically distributed local magnetic minima to eventually create degenerate quantum gases such as Bose-Einstein Condensation (BEC) and ultracold fermions. Recently, several micrometer-scale structures generating periodic magnetic potentials have been proposed for trapping cold atoms as an alternative to optical lattices [7,9]. These specifically engineered quantum devices can be created by manufacturing one dimensional and two dimensional periodic structures on a device called an Atom Chip [8].

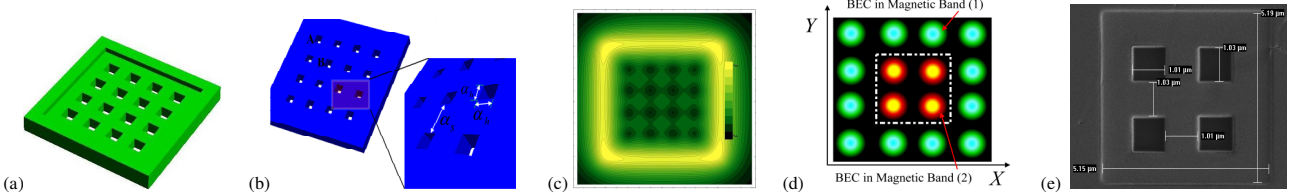

**Fig. 1.** (a) (b) Proposed nano-structured garnet-Based magnetic lattice for confinement and trapping of ultra-cold atoms. It consists of an $m \times m$ array of $n \times n$-hole blocks, where each block is surrounded by contiguous film surface areas that create magnetic "walls" ($m = 2$ and $n = 4$ in the shown structure). $\alpha_h$ represents the hole dimension while $\alpha_s$ represents the distance between two holes. The sites **A** and **B** represent two magnetic quantum wells in two different magnetic bands. (c) x-y Contour plot of the external magnetic field at the effective distance $d_{min}$ of the lattice (d) Schematic representation of BEC distribution in the asymmetric magnetic lattice (e) Fabricated double quantum well producing the similar principles of entanglement between spatially separated double magnetic quantum well.

## 2. THE DISCRETE MAGNETIC BANDS MAGNETIC LATTICE

Devices that create a one-dimensional magnetic lattice have recently been achieved, while their applications in the field of quantum information processing are still in progress [10]. We recently proposed a two-dimensional periodic magneto optical structure to produce a two dimensional magnetic lattice, detailed briefly below. The proposed structure is shown in Figures 1(a,b), where the device is created using the garnet based magneto-optical materials $Bi_2Dy_1Fe_4Ga_1O_{12}$. The un-patterned area surrounding the $n \times n$-hole matrix designed for controlling the energy levels and the magnetic field minima of the trap where it produces locally **x-y** magnetic bias fields, $B_{x\text{-}bias}$, $B_{y\text{-}bias}$. The presence of holes results in a magnetic field distribution whose non-zero local minima are located at effective $z$-distances from the holes. These minima are localized in small volumes representing the magnetic potential wells that contain a certain number of quantized energy levels for the cold atoms to occupy. In our design, we set the size of the holes $\alpha_h$ and the separation of the holes $\alpha_s$ as $\alpha_h = \alpha_s \equiv \alpha$ to simplify the mathematical derivations. For an infinite magnetic lattice the magnetic fields along x, y and z, $B_x$, $B_y$ and $B_z$, are described as

$$B_x = B_o(1 - e^{-\beta\tau})e^{-\beta|z-\tau|}\sin(\beta x) + B_{x-bias} \quad (1)$$
$$B_y = B_o(1 - e^{-\beta\tau})e^{-\beta|z-\tau|}\sin(\beta y) + B_{y-bias} \quad (2)$$
$$B_z = B_o(1 - e^{-\beta\tau})e^{-\beta|z-\tau|}[\cos(\beta x) + \cos(\beta y)] + B_{z-bias} \quad (3)$$

where $B_o = \mu_o M_z/\pi$ is the magnetic induction at the surface of the film, $\tau$ is the film thickness and a plane of symmetry is assumed at $z = 0$, $\beta = \pi/\alpha$. The magnetic periodicity of the local minima results in confined indirectly coupled magnetic quantum wells (CMQWs) as shown in Figure 2. The resulting magnetic lattice sites are spatially distributed in the *x-y* plane with discrete bands of non-zero local magnetic minima. Each magnetic band contains uni-directional adjacent and symmetrical CMQWs along the *x* and *y*-axes, while vertically along the *x-z* and *y-z* planes they create uncoupled quantum wells separated by the magnetic band gaps. The gap value and the magnetic barrier $\Delta B$ between the lattice sites can be controlled via the application of an external magnetic bias field along the negative *z*-direction. This causes a reduction in the effective magnetization $M_z$ of the thin film, and consequently affects the magnetic band coupling parameters to form directly CMQWs and allow tunneling between the lattice sites, as shown in Figure 2(h).

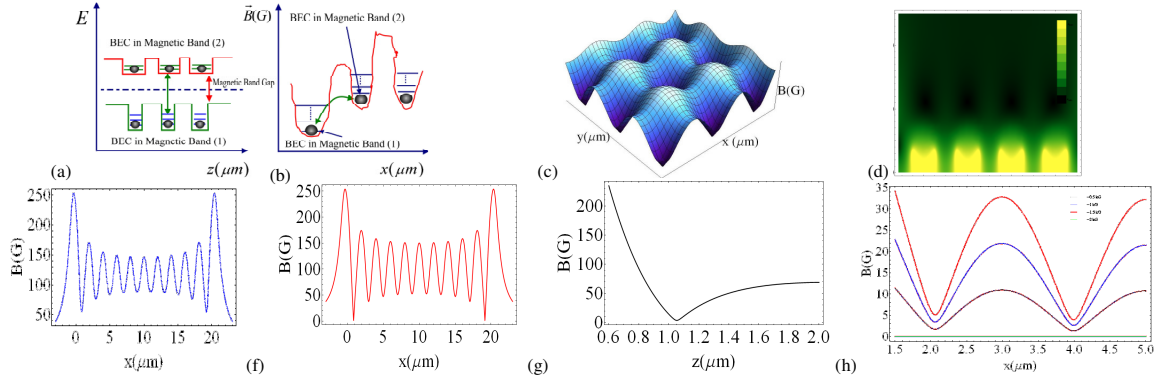

**Fig. 2.** (a) Schematic representation of the magnetic bands and (b) BEC trapped in magnetic quantum wells. (c) 3D plot of two indirectly coupled quantum wells. The magnetic barrier between the vertical sites represent the magnetic band gap which can minimized using external magnetic field $B_{z-bias}$.(d) Density plot of the magnetic field along the *z-x* plane (e) Magnetic field distribution across $10 \times 10$ magnetic lattice at the center and (f) at the edge of the trap along x-axis. (g) The non-zero magnetic minima at the effective distance from the trap surface. (h) Magnetic barrier between two sites can be minimized by applying external magnetic field $B_{z-bias}$.

### 3. LONG-RANGE ENTANGLEMENT OF MULTIPARTITE SYSTEM FORMED IN A MAGNETIC LATTICE

Ideally, long-range entanglement of several directly coupled qubits can be represented as interacting system chain of harmonic oscillators [2]. So far, entanglement of mixed states is not identified, but a method of entanglement distillation is often used to indirectly quantify the many-qubit entanglement. Our proposal offers this feature by selectively assigning a quantum pure state to a specific number of lattice sites, such as creating maximally entangled Bell state pairs, and then dynamically allowing the system to evolve into the regime of long-range entangled multipartite. As the simulation result shows, the sites are spatially well separated as required before implementing the entanglement [3]. A smaller number of maximally entangled pairs (*2×2*) can symmetrically be achieved at the center of the (*4×4*) magnetic lattice. At this stage, the system is an uncoupled harmonic oscillator chain trapped in the indirectly CMQWs. To induce long-range entanglement, tunneling between the lattice sites must be allowed which can be done by minimizing the magnetic energy band gap. The tunneling mechanism across the lattice site evolves the spatially separated qubits into interacting multipartite that formed in the indirectly CMQWs. It also allows a symmetrical distribution of BECs at the center (*m×m*) sites through successive tunneling because the CMQWs at the center are symmetrically formed. This key point allows minimum individual local control in directly entangled center qubits to the far long-range entangled qubits, where the system can be designed to allow (*n-m×n-m*) sites.

### 4. CONCLUSION

An asymmetric multi-quantum state magnetic lattice has been proposed for realising a long-range entangled multipartite system employing ultracold atoms in a Bose Einstein Condensate. Simulation results have shown that discrete magnetic bands magnetic lattice can locate a controllable long-range entanglement of a multipartite system well separated in space. The dependence of the coherence time on the long-range entanglement, which can macroscopically be identified, makes the proposed asymmetric magnetic lattice a suitable candidate to generate squeezed states for quantum information processing and communication.

### 5. REFERENCES

[1] M. B. Plenio, J. Hartley and J. Eisert, New J. Physics. **6**, 36 (2004).
[2] N. J. Cerf, G. Leuchs, E. S. Polzik, "Quantum Information with Continuous Variables of Atoms and Light", Ch **2**, p(43-62), Imperial College Press (2007).
[3] M. B. Pl, J. Eisert, J. Dreißig, and M. Cramer, Phys. Rev. Lett **94**, 060503(2005).
[4] Martin B Plenio, Fernando L Semiao, arXiv:quant-ph/0407034v2.
[5] J. Eisert, M.B. Plenio and J. Hartley, Phys. Rev. Lett. **93**, 190402 (2004).
[6] C. H. Bennett, D. P. DiVincenzo, J. A. Smolin, and W. K. Wootters, Phys. Rev. A **54**, 3824 (1996)
[7] Saeed Ghanbari, Tien D Kieu, Andrei Sidorov and Peter Hannaford, J. Phys. B, **39** 847(2006).
[8] B.V. Hall, S. Whitlock, F. Scharnberg, P. Hannaford, and A. I. Sidorov, J. Phys. B: At. Mol. Opt. Phys. **39** 27 (2006).
[9] D Jaksch D. Jaksch, C. Bruder, J. I. Cirac, C. W. Gardiner, and P. Zoller. Phys. Rev. Lett. **81** 3108 (1998).
[10] M. Singh, M. Volk, A. Akoulchine, A. Sidorov, R. McLean, and P. Hannaford, J Phys B, **41** 065301 (2008).